\documentclass{JHEP3}

\usepackage{amsmath}
\usepackage{cite}
\input epsf
\usepackage{epsfig}
\usepackage{amssymb}
\usepackage[active]{srcltx}
\usepackage{hep}
\def \be {\begin{equation}}
\def \ee {\end{equation}}
\def \bea {\begin{eqnarray}}
\def \eea {\end{eqnarray}}
\def \nn {\nonumber}

\def\bd{\begin{document}}
\def\ed{\end{document}}
\def\nn{\nonumber}
\def\bea{\begin{eqnarray}}
\def\eea{\end{eqnarray}}
\let\bm=\bibitem
\let\la=\label

\def\Label#1{\label{#1}%
  \smash{\hbox to0pt{\raise1ex\hbox{\tiny[#1]}\hss}}}
\newcommand{\bbibitem}[1]{\bibitem{#1}\marginpar{#1}}

\title{On bound states in Quantum Field Theory  }
\author{Changyong Liu\footnote{email address:
liuchangyong@nwsuaf.edu.cn}
\\
College of Science, Northwest A\&F University, Yangling, Shaanxi
712100, China}
\date{\today}
\abstract{In this paper, a new method to describe the energy spectrums of bound states in Quantum Field Theory is presented. We point out that the fundamental field and its dual soliton combine together to form bound states and the soliton corresponds to the ghost particle in our regularization scheme which takes advantage of dimensional regularization and Pauli-Villars regularization. Based on this point of view, we discuss the bound states of massive Thirring model, the positronium ($e^+e^-$) in QED and the vector meson in QCD. We also give a new way to obtain the mass of soliton (quantum soliton) from the stationary condition (gap equation). Our results agree with experimental data to high precision. We argue that the hypothetic $X_{17}$ particle in decay of $\rm{{}^8 Be}$ and $\rm{{}^4 He}$ is a soliton. For vector meson, we find the squared masses of $\rho$ resonances are $m^2(n)\sim (an^{1/3}-b)^2$ which coincide well with experiments. }

\begin{document}

%%%%%%%%%%%%
\section{Introduction}
The correlation function and S-matrix are the basic building blocks of Quantum Field Theory (QFT). These quantities can be calculated perturbatively at weak coupling. The perturbative QFT have achieved great successes \cite{Witten:2003nn}, but it is insufficient to solve the non-perturbative aspects
 of QFT. We need new methods to study the
non-perturbative problem especially the bound state which are important objects in quantum theory and have been widely studied. There are many methods aim to solve it completely, e.g. the Bethe-Salpeter equation \cite{Salpeter:1951sz} and Lattice QCD \cite{Wilson:1974sk}. The Bethe-Salpeter equation is difficult to solve. Lattice calculations give us useful numerical results, but we still need to understand many important phenomena from analytical aspects. In this paper we study the bound states from the analytical structure of the correlation function. The general representations of the vacuum expectation value of two Heisenberg operators are given by the $\rm{K\ddot{a}ll\acute{e}n-Lehmann}$ spectral representation \cite{Kallen:1952zz}. We give the $\rm{K\ddot{a}ll\acute{e}n-Lehmann}$ form for the vacuum expectation value of the time-ordered product of two vector field which
can be obtained from that of the ordinary product of the two operators, the Wightman function:
\be
\langle\Omega| A_{\mu}(x)A_{\nu}(y)|\Omega\rangle=\sum_n\langle\Omega|A_{\mu}(x)|n\rangle
\langle n| A_{\nu}(y)|\Omega\rangle,\nonumber
\ee
where the complete set of states $\{|n\rangle\}$ has been used. Based on general assumptions about invariance and the spectrum, the
expression of the Wightman function is
\be
\langle\Omega| A_{\mu}(x)A_{\nu}(y)|\Omega\rangle=i\int_0^{\infty} dm^2 \rho_{\mu\nu}(m^2)\Delta^{(+)}(x-y;m^2).\nonumber
\ee
The theory's content is determined by the spectral density function $\rho_{\mu\nu}$. The exact Feynman propagator for the gauge field in the $\rm{K\ddot{a}ll\acute{e}n-Lehmann}$ spectral representation is given by
\be \label{KL}
\langle\Omega| T(A_{\mu}(x)A_{\nu}(y))|\Omega\rangle=\frac{1}{2}\int_0^{\infty} dm^2 \rho_{\mu\nu}(m^2)\Delta_F(x-y;m^2).
\ee
If a bound state exists in the theory, the pole of the Fourier transform of equation (\ref{KL}) gives the mass of the bound state. Based on the perturbatively calculation, we fail to obtain the mass spectrums of bound states from the vacuum expectation value of two Heisenberg operators. We find that the spectrums of bound states can be obtained by taking into account the soliton contribution. The solitons play an important role in QFT and be related to fundamental particles. Dirac \cite{Dirac:1931kp} argued that in quantum mechanics the unobservability of phase permits singularities which manifest themselves as sources of magnetic fields and the electric charge $q$ and magnetic charge $g$ satisfies the quantization condition
\be \label{charge}
qg=\frac{n}{2}
\ee
with $n$ an arbitrary integer. The Dirac quantization condition is a topological property. The magnetic monopoles necessarily arise as soliton in gauge theories the 't Hooft-Polyakov monopole \cite{tHooft:1974kcl}. The free monopole has not yet been discovered due to the mass $M\sim \frac{M_{\rm{w}}}{\alpha}\sim 137M_{\rm{w}}$, where $M_{\rm{w}}$ is the mass of the intermediate vector boson of the weak interaction. There is another particle known as the dyon carrying electric and magnetic charge. The mass of a dyon of magnetic number $n_e$ and electric number $n_m$ is \cite{Prasad:1975kr}
\be
M\geq \sqrt{2}|Z|
\ee
with complex charge
\be \label{charge1}
Z=v(n_e+i\frac{1}{\alpha}n_m).
\ee
Where $v$ is the Higgs expectation value and $\alpha$ is $\alpha=\frac{g^2}{4\pi}$. In bound states, we find that the mass of fundamental field and its dual soliton satisfies a new relation similar to the Dirac quantization condition (\ref{charge}). From this relation, we can study the energy eigenvalues of the bound states and the mass of soliton. This also indicates that the fundamental field and its dual soliton combine together to form bound states. Besides this, we give a clue for physical regularization in QFT.

The paper is organized as follows. In Section 2, we give a general framework of our method. Based on a new regularization scheme, we obtain a formulae
of energy eigenvalues of the bound states. We argue that the  mass of the fictitious particle in Pauli-Villars regularization is the same as the dual soliton mass. In Section 3, we discuss the bound states of massive Thirring model to show our method can obtain the right results. In Section 4, we discuss the positronium ($e^+e^-$) in QED. We give a new way to obtain the mass of soliton from the stationary condition (gap equation) by analytic continuation. In Section 5, we consider vector meson in QCD. We end with the conclusions. \\

%%%%%%%%%%%%%%%%%%
\section{A general framework }
The energy eigenvalues of bound states are the poles of propagator. The general form of off shell propagators  $G_2(p^2)$ by the chain approximation in momentum sapce have a factor \cite{Peskin:1995ev}
\be
G_2(q^2)\propto\frac{1}{1-\Pi(q^2,m)},
\ee
where the $\Pi(q^2,m)$ is obtained from loop diagram and contains ultraviolet (UV) divergence. The UV divergence can be regulated by the Pauli-Villars (PV) \cite{Pauli:1949zm} regularization and dimensional regularization\cite{tHooft:1972tcz}. The PV regularization requires that for each particle of mass $m$ a new unphysical ghost particle of mass $\widetilde{m}$ be added with either the wrong statistics or wrong-sign kinetic term. The PV regularization is complicated because it involves introducing, in a gauge invariant manner, sets of heavy fields. The dimensional regularization respects gauge invariance and can regulate IR or UV divergences, but it has the disadvantage of being unphysical. We use a new regularization scheme which takes advantage of dimensional regularization and Pauli-Villars regularization. We first calculate the Feynman diagrams in $d=n-2\varepsilon$ dimensions, then we add a ghost particle to cancel the UV divergence, where the ghost particle is determined by the bound states. The new regularization scheme will be physical one without any fictitious particle. Applying the new regularization scheme in $G_2(p^2)$, it modifies the propagator $G_2(p^2)$ as follows:
\be
G_2(q^2)\propto\frac{1}{1-\Pi(q^2,m)}\rightarrow \widetilde{G}_2(q^2)\propto\frac{1}{1-(\Pi(q^2,m)-\Pi(q^2,\widetilde{m}))},
\ee
Where the $\Pi(q^2,m)$ is calculated in $d=n-2\varepsilon$ dimensions. Then the energy eigenvalues of the bound states are the solutions of the equation
\be \label{sol}
\lim_{\varepsilon\rightarrow 0}[1-(\Pi(q^2,m)-\Pi(q^2,\widetilde{m}))]=0.
\ee
To get the excited spectrum of bound states, we need to know the fictitious particle mass $\widetilde{m}$.
We find that the mass of the fictitious particle is the same as the dual soliton mass:

\be
\boxed {{\rm ghost\ particle}}\iff \boxed{{\rm soliton\ solution}},
\ee
then we have the relation

\be
\Pi^{{\rm soliton}}(q^2,\widetilde{m})=-\Pi(q^2,\widetilde{m}).
\ee
From this we say that the solition contributions give a natural physical UV subtraction. According to the equation (\ref{sol}), the soliton mass (quantum soliton) can be obtained by the energy eigenvalues of bound states.

\begin{figure}[h]
\center
\includegraphics[width=0.3\textwidth]{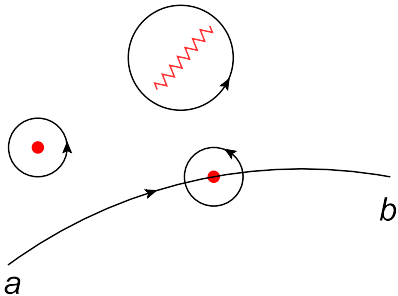}
\caption{\label{2} A smooth contour $C[a,b]$ in complex plane staring from $a$ to $b$. The red dots and wave line denote the poles and branch cut  of function $f(z)$ separately.  }
\end{figure}

In order to study the bound states, we define the integral of a complex function $f(z)$ along a smooth contour $C[a,b]$ in complex plane. Suppose the function $f(z)$ have poles or branch cuts (FIG. \ref{2}), then the integral of $f(z)$ along the contour $C[a,b]$ can be expressed as
\par

\bea
\label{fz}
\int_{C[a,b]}f(z)dz=\int_a^bf(z)dz+\sum n_i \oint_{C_i}f(z)dz.
\eea

Where the $C_i$ is a closed curve circling the pole or branch cut. The $\int_a^bf(z)dz$ takes value in main single-valued branch range. The winding number $n_i\in Z$ is the
contour circling $n_i$ times around the pole or branch cut. For example, we obtain the following integral

\be
\int_{C[-1,1]}\frac{1}{z}dz=\int_{-1}^1\frac{1}{z}dz+2\pi n i=\log(-1)+2\pi n i=\pi i+2\pi n i,
\ee
where $n\in Z$ and ${\rm Im}\int_{-1}^1\frac{1}{z}dz\in (-\pi, \pi]$.

\section{The bound states of massive Thirring model}
\par
We first discuss the ground state of massive Thirring model \cite{Thirring:1958in} which is a theory of a fermion field $\psi$ in (1+1)-dimensional space-time, with the Lagrangian density
\be \label{Thirring}
\mathcal{L}=\bar{\psi}(i\gamma^{\mu}\partial_{\mu}-M)\psi-\frac{g}{2}(\bar{\psi}\gamma^{\mu}\psi)^2.
\ee
Coleman \cite{Coleman:1974bu} proved that the zero-charge sector of the massive Thirring model is equivalent to the sine-Gordon theory with the Lagrangian density
\be
\mathcal{L}=\frac{1}{2}\partial_{\mu}\phi\partial^{\mu}\phi+\frac{m^4}{\lambda}[\cos(\frac{\sqrt{\lambda}}{m}\phi)-1].
\ee
The sine-Gordon kink mass and the breather mass are given by \cite{Dashen:1974ci}
\bea \label{breather}
&&M_{kink}=\frac{8m}{\gamma'}+O(\lambda/m),\nonumber \\
&&M_n=\frac{16m}{\gamma'}\sin(\frac{n\gamma'}{16}),  \qquad n=1,2,\ldots,<\frac{8\pi}{\gamma'},
\eea
where $\gamma'=\frac{\lambda}{m^2}(1-\frac{\lambda}{8\pi m^2})^{-1}$. To examine the spectrum for $\lambda$ just below $4\pi m^2$, we write
\be
\frac{\lambda}{4\pi m^2}=\frac{1}{1+g/\pi}.
\ee
The energy of the lowest breather mode can be written as
\be \label{breather}
M_1=M_{{\rm kink}}[2-g^2+\frac{4g^3}{\pi}+O[g]^4].
\ee
The weak-coupling $\frac{\lambda}{m^2}\rightarrow 0$ limit of the sine-Gordon theory would correspond to the infinity strong-coupling $g\rightarrow \infty$ limit of the massive Thirring model, while the $g\rightarrow 0$ weak-coupling limit of the latter would
correspond to the strong coupling limit $\frac{\lambda}{m^2}\rightarrow 4\pi$. To check the correspondences, we study the bound states of massive Thirring model
in the weak-coupling limit $g\rightarrow 0$. We review some results on the bound states of massive Thirring model which have been discussed in \cite{Aoyama:1977yf} by the chain approximation. The Fourier component of the propagator for a scalar state is
\bea
\Delta(k)=\frac{1}{i(2\pi)^2}\int d^2x<0\mid T\bar{\psi}(x)\psi(x)\cdot\bar{\psi}(y)\psi(y)\mid0>e^{-ik(x-y)}
=\frac{\Pi_s(k^2)}{1+\frac{g}{2}\Pi_s(k^2)}.
\eea
Where the $\Pi_s(k^2)$ is given by
\bea
\Pi_s(k^2)=\frac{1}{i}\int\frac{d^2 q}{(2\pi)^2}{\rm Tr}\frac{1}{((q^{\mu}+k^{\mu}/2)\gamma_{\mu}-M)((q^{\nu}+k^{\nu}/2)\gamma_{\nu}-M)}.
\eea
The poles of the propagator represent the energy eigenvalues of a scalar bound states
\bea \label{ggg}
1+\frac{g_0}{2}\Pi_s(k^2)=0. \nonumber
\eea
The $\Pi_s(k^2)$ has ultraviolet divergence. To remove the divergence, we need renormalization. One method is to make the
replacement $\Pi_s(k^2)\rightarrow\widehat{\Pi}_s(k^2)=\Pi_s(k^2)-\Pi_s(0)$ and obtain
\bea
\widehat{\Pi}_s(k^2)&=&\frac{1}{2\pi}\int_0^1 dx\ln(1-x(1-x)t^2)\nonumber \\
&=&\frac{1}{\pi}[\sqrt{\frac{4}{t^2}-1}\arctan(\frac{t}{\sqrt{4-t^2}})-1],
\eea
\par
\begin{figure}[h]
\center
\includegraphics[width=0.3\textwidth]{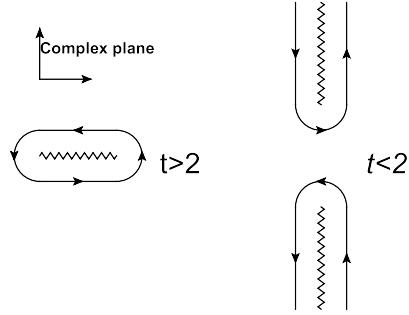}
\caption{\label{cuts4} The branch cuts of function $f(x)=\ln(1-x(1-x)t^2)$ with different $t$. The contour integral along the semi-infinite branch cut is infinite for $t<2$. }
\end{figure}
\begin{figure}[h]
\center
\includegraphics[width=0.5\textwidth]{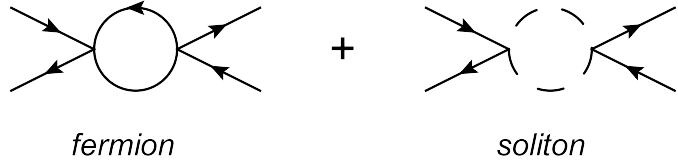}
\caption{\label{soliton} The bound states have two main contributions. The soliton mass $\widetilde{M}$ is $\widetilde{M}\approx \frac{2M}{g}$. }
\end{figure}


where we have defined the new variable $t=\sqrt{\frac{k^2}{M^2}}$. If we only consider the fermion contribution, the energy eigenvalues equation (\ref{ggg}) have no solutions of bound states \cite{Aoyama:1977yf}. The reason is that the contour integral along the semi-infinite branch cut of the function $f(x)=\ln(1-x(1-x)t^2)$ is infinite for $t<2$ (FIG. \ref{cuts4}). We point out that we need to take into account the soliton contribution and apply the formulae (\ref{fz}) to obtain the energy eigenvalues of bound states (FIG. \ref{soliton}).

Then the energy eigenvalues of the bound states are solutions of equation
\be \label{ggg1}
1+\frac{g_0}{2}\Pi_s(t^2)+\frac{g_0}{2}\Pi_s^{{\rm soliton}}(\widetilde{t}^2)=1+\frac{g_0}{4\pi}\int_0^1 dx\ln(1-x(1-x)t^2)
-\frac{g_0}{4\pi}\int_0^1 dx\ln(1-x(1-x)\widetilde{t}^2)
+\frac{g_0}{4\pi}\ln\frac{M^2}{\widetilde{M}^2}=0.
\ee
Where the $\Pi_s^{{\rm soliton}}(\widetilde{t}^2)$ is the soliton contribution and the new variable $\widetilde{t}$ is defined as $\widetilde{t}=\sqrt{\frac{k^2}{\widetilde{M}^2}}$. The term $\frac{g_0}{4\pi}\ln\frac{M^2}{\widetilde{M}^2}$ has the property that is $\frac{g_0}{4\pi}\ln\frac{M^2}{\widetilde{M}^2}\rightarrow 0$ for small $g_0$, we defined the renormalized coupling constant as following
\be
g=\frac{g_0}{1+\frac{g_0}{4\pi}\ln\frac{M^2}{\widetilde{M}^2}}.
\ee

Then the equation (\ref{ggg1}) becomes
\bea \label{ggg2}
 1+\frac{g}{4\pi}\int_{C[0,1]} dx\ln(1-x(1-x)t^2)
-\frac{g}{4\pi}\int_{C[0,1]} dx\ln(1-x(1-x)\widetilde{t}^2)=0.
\eea

\par
\begin{figure}[h]
\center
\includegraphics[width=0.2\textwidth]{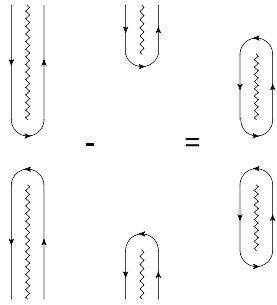}
\caption{\label{cuts} After taking into account the soliton contribution, the contour integrals along the branch cuts become finite. }
\end{figure}

 The soliton mass $\widetilde{M}$ of massive Thirring model can be obtained by the inverse scattering method  \cite{Orfanidis:1975ut}.
From the Lagrangian density (\ref{Thirring}), we can obtain the equation of motion
\bea
(i\gamma^{\mu}\partial_{\mu}-M)\psi-g(\bar{\psi}\gamma^{\mu}\psi)\gamma_{\mu}\psi=0.
\eea
The equation have a classical solution which behaves as an elementary particle with mass $\widetilde{M}=\frac{2M}{g}$. The quantum corrections to the soliton mass can be obtained similar to the sine-Gordon kink mass which is \cite{Orfanidis:1975ut}
\bea \label{masssoliton}
\frac{\widetilde{M}}{M}=\frac{2}{g}-\frac{1}{\pi}+\frac{3g}{4}+\frac{17g^2}{24\pi}+O[g]^3.
\eea

According to the formulae (\ref{fz}) and the contour integral along the branch cuts (FIG. \ref{cuts}), the equation (\ref{ggg2}) becomes
\bea \label{engy}
1&+&\frac{g}{2\pi}[\sqrt{\frac{4}{t^2}-1}\arctan(\frac{t}{\sqrt{4-t^2}})-1]-
\frac{g}{2\pi}[\sqrt{\frac{4}{\widetilde{t}^2}-1}\arctan(\frac{\widetilde{t}}{\sqrt{4-\widetilde{t}^2}})-1]\nonumber \\
&-&\frac{g}{2}\sqrt{\frac{4}{\widetilde{t}^2}-1}+\frac{g}{2}\sqrt{\frac{4}{t^2}-1}=0.
\eea

From the equations (\ref{masssoliton}) and (\ref{engy}), we obtain the ground state mass
\be
M_{bound}=M[2-g^2+\frac{4g^3}{\pi}+O[g]^4].
\ee
Which is consistent with the lowest breather mass of sine-Gordon model given by (\ref{breather}).

\section{Positronium ($e^+e^-$) systems in QED}
 \par
Then we study the positronium ($e^+e^-$) bound states. We review some results of QED which can be found in famous papers and text books \cite{Pauli:1949zm,Schwinger:1951nm,Peskin:1995ev}. We define $i\Pi^{\mu\nu}(q)$ to be the sum of all 1-particle-irreducible (1PI) insertions into the photon propagator. The $\Pi^{\mu\nu}(q)$ has the tensor
structure:
\be
\Pi^{\mu\nu}(q)=(q^2g^{\mu\nu}-q^{\mu}q^{\nu})\Pi(q^2).\nonumber
\ee
The $\alpha$ ($\alpha=\frac{e^2}{4\pi}$) contribution to $i\Pi^{\mu\nu}(q)$ is $i\Pi^{\mu\nu}_2(q)=i(q^2g^{\mu\nu}-q^{\mu}q^{\nu})\Pi_2(q^2)$ coming from the following electron loop
\par
\begin{figure}[h]
\center
\includegraphics[width=0.2\textwidth]{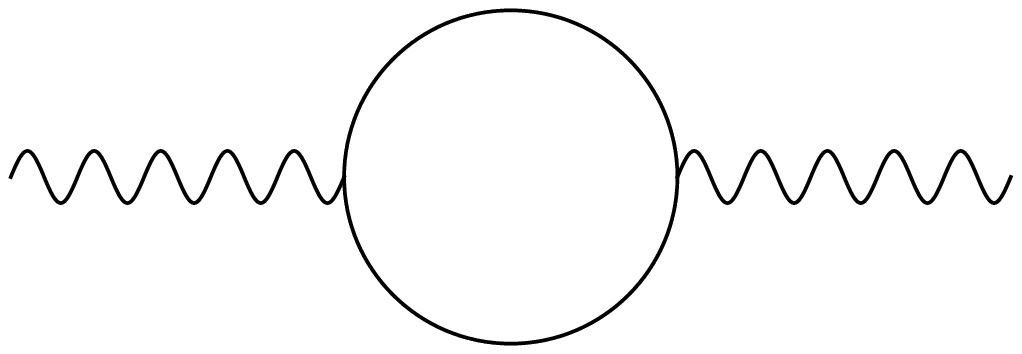}

\end{figure}
 and
\bea
 \Pi_2(q^2)&=&-8e^2\int_{0}^{1} dx x(1-x)\int\frac{d^4l_E}{(2\pi)^4}\frac{1}{(l_E^2+\Delta)^2}\nonumber\\
 &=&-\frac{2\alpha}{\pi}\int_{0}^{1} dx x(1-x)\int_0^{\infty}\frac{ydy}{(y+\Delta)^2},
\eea
where $\Delta=m_e^2-x(1-x)q^2$. The excited spectrums are the poles of the two-point function
\bea
G_{\mu\nu}(q^2)=\frac{-ig_{\mu\nu}}{q^2(1-\Pi(q^2))}.\nonumber
\eea

The two-point function has ultraviolet divergence. We choose our new scheme to regulate the ultraviolet divergence
\bea
\widetilde{G}_{\mu\nu}(q^2)=\frac{-ig_{\mu\nu}}{q^2(1-[\Pi(q^2,m_e)-\Pi(q^2,\widetilde{m}_e)])}.\nonumber
\eea
Where the $\widetilde{m}_e$ is the mass of the soliton in QED that is $\Pi^{{\rm soliton}}_2(q^2,\widetilde{m}_e)=-\Pi(q^2,\widetilde{m}_e)$.

\begin{figure}[h]
\center
\includegraphics[width=0.5\textwidth]{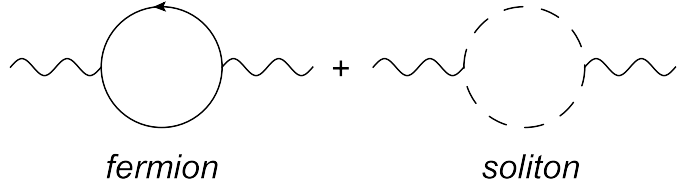}
\caption{\label{fsloop} The bound states have two main contributions. }
\end{figure}

We consider the positronium ($e^+e^-$) bound states which correspond to the case $t<2$. To simplify our discussion, we define the $\widehat{\Pi}_2(q^2)$ as
\bea \label{pi}
\widehat{\Pi}_2(q^2)&\equiv& \Pi_2(q^2,m_e)-\Pi_2(0,m_e)\nonumber\\
 &=&-\frac{2\alpha}{\pi}\int_{C[0,1]}dxx(1-x)\log(\frac{1}{1-x(1-x)t^2})\nonumber.
\eea
The $C[0,1]$ represents a smooth contour staring from $0$ to $1$. We defined the new variable $t=\sqrt{\frac{q^2}{m_e^2}}$. Similar to the massive Thirring model, the contour integral along the semi-infinite branch cut of the function $f(x)=\ln(1-x(1-x)t^2)$ is infinite for $t<2$ (FIG. \ref{cuts4}). We need to take into account the soliton contribution $\Pi^{{\rm soliton}}_2(q^2,\widetilde{m}_e)$ (FIG. \ref{fsloop}).
We define the $\widehat{\Pi}^{{\rm soliton}}_2(q^2)$ as

\bea \label{pi1}
\widehat{\Pi}^{{\rm soliton}}_2(q^2)&\equiv& -(\Pi_2(q^2,\widetilde{m}_e)-\Pi_2(0,\widetilde{m}_e))\nonumber\\
 &=&\frac{2\alpha}{\pi}\int_{C[0,1]}dxx(1-x)\log(\frac{\widetilde{m}_e^2}{\widetilde{m}_e^2-x(1-x)q^2})\nonumber.
 \eea
 \begin{figure}[h]
\center
\includegraphics[width=0.4\textwidth]{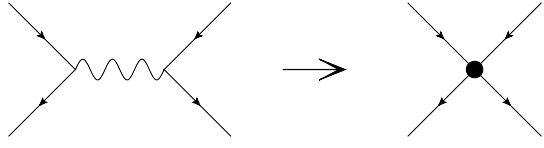}
\caption{\label{four}. The effective theory of bound states is 4-Fermi theory with coupling constant $G_F\approx\frac{e^2}{2m^2_e}$. }
\end{figure}

The mass of soliton in QED is difficult to solve. We need a new way to calculate it. The photon in bound states becomes massive, so we consider the 4-Fermi effective theory (FIG. \ref{four}) with Lagrangian density
\be \label{Thirring1}
\mathcal{L}(\psi,\bar{\psi})=\bar{\psi}(i\gamma^{\mu}\partial_{\mu}-m)\psi+\frac{G_F}{2}(\bar{\psi}\psi)^2,
\ee
 which is known as the massive Gross-Neveu (GN) model \cite{Gross:1974jv}. Where the coupling constant $G_F$ is $G_F\approx\frac{e^2}{2m^2_e}$. The Lagrangian (\ref{Thirring1}) can be re-written by using a scalar auxiliary field $\sigma(x)$ :
\be \label{Thirring2}
\mathcal{L}(\psi,\bar{\psi},\sigma)=\bar{\psi}(i\gamma^{\mu}\partial_{\mu}-m)\psi-\sigma\bar{\psi}\psi-\frac{1}{2G_F}\sigma^2.
\ee
The (\ref{Thirring2}) is classically equivalent to (\ref{Thirring1}) by the equation of motion for $\sigma(x)$.
The Lagrangian (\ref{Thirring2}) is purely quadratic in the fermion field we can integrate directly
\be
\int\mathcal{D}\bar{\psi}\mathcal{D}\psi\mathcal{D}\sigma\exp[i\int d^dx\mathcal{L}(\psi,\bar{\psi},\sigma)]=
\int\mathcal{D}\sigma\exp[i\int d^dx(-\frac{1}{2G_F}\sigma^2)+{\rm Tr}(\log(i\gamma^{\mu}\partial_{\mu}-m-\sigma))].
\ee
Then the effective theory can be calculated by the method of steepest descents. The vacuum expectation value $M$ of the field $\sigma(x)$
is the solution of the stationary condition (gap equation)
\be \label{dsss}
-\frac{M}{G_F}+\int\frac{d^dp}{(2\pi)^d}{\rm Tr}(\frac{i}{\slashed{p}-m-M})=0.
\ee
We find a new method to deal with the gap equation (\ref{dsss}) by using the formulae (\ref{fz}). From the gap equation (\ref{dsss}), we can obtain the mass
of soliton. Let us briefly recall the results of GN model with $N$ Dirac fermions in $d=1+1$ dimensions. The solitons in the massless ($m=0$) GN model are Callan-Coleman-
Gross-Zee (CCGZ) kink\cite{CCGZ} and DHN saddle point configurations (fermion bag solitons)\cite{Dashen:1975xh} in the large $N$ limit.
The ground state is fixed by the gap equation
\be \label{dsssn}
-\frac{M_0}{G_F}+N\int\frac{d^dp}{(2\pi)^d}{\rm Tr}(\frac{i}{\slashed{p}-M_0})=0,
\ee
in which $M=M_0$ is constant.
The dynamically generated mass of small fluctuations of the Dirac fields
around this vacuum is
\be \label{dsssnmu}
m=M_0=\widetilde{\mu}\exp^{1-\frac{\pi}{NG_F(\widetilde{\mu})}},
\ee
where the $\widetilde{\mu}$ is an the renormalization scale.
The mass of CCGZ kink is
\be \label{dsssncc}
M_{\rm{kink}}=\frac{mN}{NG_F}=\frac{m}{G_F}=\frac{mN}{\pi},
\ee
where the $NG_F=\pi$ by choosing the renormalization point at $\widetilde{\mu}=m$. The mass of fermion bag soliton is
\be
M_{\rm{soliton}}(n)=\frac{2mN}{\pi}\sin(\frac{\pi n}{2N}),
\ee
which is the bound state of kinks.
The solitons in the $1 + 1$ dimensional massive
generalization of the GN model was carried in\cite{Feinberg:1996kr} with mass
\be \label{dsssnmummm}
M_{\rm{soliton}}(v)=Nm(\frac{2}{\pi}\sin\theta+\gamma\log\frac{1+\sin\theta}{1-\sin\theta}).
\ee
In massive ($m\neq 0$) GN model, the CCGZ kink becomes infinitely massive and disappears by classical analysis. We argue that there is quantum effect to allow the
CCGZ kink exist, which is quantum soliton \cite{Davies:2019wym}.

  We now use a new method to find mass of soliton from gap equation (\ref{dsss}). In our approach, we adopt the new regularization scheme. We let  $d=2-2\varepsilon$ and introduce a parameter $\mu$ with dimensions of mass. Unlike the $\widetilde{\mu}$ in (\ref{dsssnmu}) which is an the renormalization scale, the $\mu$ is determined by the the gap equation (\ref{dsss}). If we consider mass $m$ to be the physical mass of fermion, then the $M$ is pure imaginary. Similar to the imaginary part of complex charge $Z$ (\ref{charge1}), the $M$ connects with the soliton mass. We assume that the $M$ is $M=i\widetilde{m}$, then the equation (\ref{dsss}) becomes
\bea
-\frac{i\widetilde{m}}{G_F}&=&-2(m+i\widetilde{m})i\mu^{2\varepsilon}\int\frac{d^dp}{(2\pi)^d}(\frac{1}{p^2-(m+i\widetilde{m})^2})
\nonumber\\
&=&\frac{1}{2\pi}(m+i\widetilde{m})(\frac{\Gamma(1-\frac{d}{2})}{((\frac{m+i\widetilde{m}}{\mu})^2)
^{1-\frac{d}{2}}}
-2\pi \widetilde{n} i).
\eea
Because the $\log(x)$ function is multi-valued function, we have the $2\pi \widetilde{n} i$ term with $\widetilde{n}\in Z$ in the above expression. The solutions of the equation have two branches:
$$\widetilde{m}^2+m^2=
\begin{cases}
\frac{\widetilde{m}m}{\widetilde{n}G_F}& {\rm for}\:\: \mu^2<0; \\
\frac{\widetilde{m}m}{(\widetilde{n}-1/2)G_F}&{\rm for} \:\: \mu^2>0.
\end{cases}$$

Then we obtain
$$\widetilde{m}\approx
\begin{cases}
\frac{m}{\widetilde{n} G_F}& {\rm CCGZ \:\: kink \:\:( \:quantum\:\: soliton \:)}; \\
\frac{m}{(\widetilde{n}-1/2) G_F}&{\rm fermion \:\:bag \:\:soliton}.
\end{cases}$$

The solutions are the quantum solitons \cite{Davies:2019wym} which also include the classical configurations. When we let $\widetilde{n}=1$, we repeat the mass of CCGZ kink (\ref{dsssncc}) and fermion bag soliton (\ref{dsssnmummm}) at kink-antikink threshold in leading order term.

We now consider the soliton mass of massive GN model in four dimensions. As the same as  massive GN model in the $1+1$ dimensions, we let $d=4-2\varepsilon$ and $M=i\widetilde{m}$. Then the gap equation (\ref{dsss}) becomes
\bea
-\frac{i\widetilde{m}}{G_F}&=&-4(m+i\widetilde{m})i\mu^{2\varepsilon}\int\frac{d^dp}{(2\pi)^d}(\frac{1}{p^2-(m+i\widetilde{m})^2})
\nonumber\\
&=&\frac{(m+i\widetilde{m})^3}{4\pi^2}(\frac{\Gamma(1-\frac{d}{2})}{((\frac{m+i\widetilde{m}}{\mu})^2)^{2-\frac{d}{2}}}+2\pi \widetilde{n} i).
\eea
From this, we find two branch solutions

$$\widetilde{m}\approx
\begin{cases}
\sqrt[3]{\frac{6\pi m}{\widetilde{n}G_F}}\approx m\sqrt[3]{\frac{3}{\widetilde{n}\alpha}} & {\rm for}\:\: \mu^2<0; \\
\sqrt[3]{\frac{6\pi m}{(\widetilde{n}-1/2)G_F}}\approx m\sqrt[3]{\frac{3}{(\widetilde{n}-1/2)\alpha}}&{\rm for} \:\: \mu^2>0.
\end{cases}$$

The soliton solutions with mass $m\sqrt[3]{\frac{3}{\widetilde{n}\alpha}}$ are similar to the CCGZ kinks in two dimensions. This is the leading term of the soliton mass $\widetilde{m}_e$ that is $\widetilde{m}_e\sim m_e\sqrt[3]{\frac{3}{n\alpha}}$ in QED. To obtain the correct bound states, we find the $\widetilde{m}_e$ can be expressed as
\bea \label{mm}
\frac{\widetilde{m}_e^2(\alpha,n,a)}{m_e^2}\equiv\Gamma(\alpha,n,a)&=&-\frac{\left(\alpha ^2-8 n^2\right)^2}{n^4}+\frac{\left(\alpha ^3-8 \alpha  n^2\right)^4}{n^4\Upsilon(\alpha,n,a)^{1/3}}+
\frac{\Upsilon(\alpha,n,a)^{1/3}}{n^4\alpha^4}\nonumber\\
&=&\frac{\left(\frac{3}{2}\right)^{2/3} \sqrt[3]{\frac{a^2}{n^2}}}{\alpha ^{2/3}}-1+\frac{\left(\frac{2}{3}\right)^{2/3} \alpha ^{2/3}}{\sqrt[3]{\frac{a^2}{n^2}}}+O[\alpha ]^{4/3}\nonumber\\
&=&\Omega(n\alpha,a)+O[\alpha ]^{4/3}.
\eea
Where the $\Upsilon(\alpha,n,a)$ is defined as
\bea
\Upsilon(\alpha,n,a)\equiv \alpha ^{12} \left(\alpha ^2-8 n^2\right)^6-2 a^2 \alpha ^{10} \left(\alpha ^2-16 n^2\right) \left(\alpha ^4+96 n^4-16 \alpha ^2 n^2\right)^2+4096n^{12}  \nonumber\\
\sqrt{\alpha ^{16} \left(\left(-2 a^2 \left(2-\frac{\alpha ^2}{4 n^2}\right)^6+24 a^2 \left(2-\frac{\alpha ^2}{4 n^2}\right)^2+32 a^2+\alpha ^4 \left(2-\frac{\alpha ^2}{4 n^2}\right)^6\right)^2-\alpha ^8 \left(2-\frac{\alpha ^2}{4 n^2}\right)^{12}\right)}\nonumber. \\
\eea
The function $\Omega(x,a)$ is defined as
\bea
\Omega(x,a)=(\frac{3a}{2x})^{2/3}+(\frac{2x}{3a})^{2/3}-1.
\eea
The constant $a$ is $a\approx 2.00408$ for positronium ($e^+e^-$) bound states.

\begin{figure}[h]
\center
\includegraphics[width=0.4\textwidth]{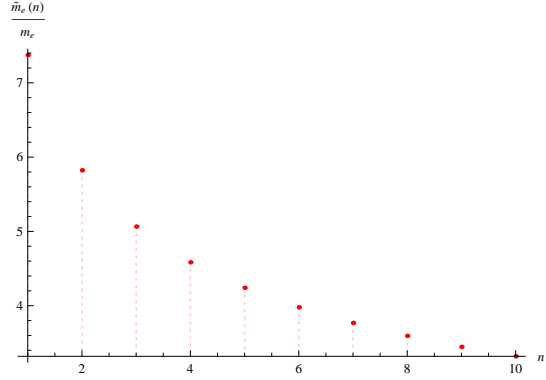}
\caption{\label{me} The curve of function $\frac{\widetilde{m}_e(n)}{m_e}$ with $a\approx 2.00408$ and $\alpha^{-1}= 137.03599913$. The soliton mass $\widetilde{m}_e$ corresponding to the ground state is $\widetilde{m}_e(1)\approx7.4 m_e$. }
\end{figure}

To illustrate the property of the $\widetilde{m}_e$, we put the
curve of function $\frac{\widetilde{m}_e(n)}{m_e}$ in FIG. \ref{me}. The soliton mass $\widetilde{m}_e$ corresponding to the ground state is $\widetilde{m}_e(1)\approx7.4 m_e$. Similar to the above discussion, we present a soliton interpretation to the hypothetic $X_{17}$ particle which has mass $m_{x_{17}}\approx 32.681m_e$\cite{Krasznahorkay:2015iga}. Anomaly in decay of $\rm{{}^8 Be}$ and $\rm{{}^4 He}$ is interpreted as a signature of a decay via emission of neutral boson $X_{17}$. To discuss the $X_{17}$,  we need to study the photon interacting with up quarks. The effective coupling constant $\alpha_u$ is $\alpha_u=N_cQ_f^2\alpha$, where $N_c=3$ is color factor and $Q_f=2/3$ is the electric charge of the up quark. Using the same formulae as (\ref{mm}), we obtain the soliton mass $\widetilde{m}_u\approx31.4093m_e\approx m_{x_{17}}$. The almost same mass indicate that the hypothetic $X_{17}$ particle is the soliton (quantum soliton).

\par
\begin{figure}[h]
\center
\includegraphics[width=0.2\textwidth]{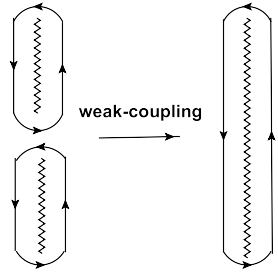}
\caption{\label{cuts2} Two branch cuts can be approximated as one single branch cut in weak-coupling region.  }
\end{figure}

According to the formulae (\ref{fz}) and the contour integral along the branch cuts (FIG. \ref{cuts}, FIG. \ref{cuts2}), the energy spectrum of positronium is solution of the equation
\bea
\label{qede}
&&1-\Pi_2(t^2)-\Pi^{{\rm soliton}}_2(t^2)\approx 1-\widehat{\Pi}_2(t^2)-\widehat{\Pi}^{{\rm soliton}}_2(t^2)\nonumber\\
&&\approx 1+\frac{\alpha(t\sqrt{4-t^2}(12+5t^2)+6(-8-2t^2+t^4)\arctan[\frac{t}{\sqrt{4-t^2}}])}
{9\pi t^3\sqrt{4-t^2}}-4n\alpha i \int_{\widetilde{x}_1}^{\widetilde{x}_2} dx x(1-x)\nonumber\\
&&= 1+\frac{\alpha(t\sqrt{4-t^2}(12+5t^2)+6(-8-2t^2+t^4)\arctan[\frac{t}{\sqrt{4-t^2}}])}
{9\pi t^3\sqrt{4-t^2}}-\frac{2an}{3\alpha}\sqrt{\frac{4}{t^2}-1}(1+\frac{2}{t^2})=0.
\eea
The $\widetilde{x}_{1,2}=\frac{1}{2}\pm\frac{1}{2}\sqrt{1-\frac{4\Gamma(\alpha,n,a)}{t^2}}$ are the solutions of equation $\Delta=\widetilde{m}_e^2-x(1-x)q^2=0$. The solution is consistent with the known result of positronium to high precision. To illustrate this, we compare the $t-2$ of the solution of (\ref{qede}) with the result of (\ref{qedd})
in Table \ref{tabi1} with $\alpha^{-1}= 137.03599913$. The excited spectrum of positronium to order $\alpha^2$ is \cite{Itzykson:1980rh}
\be
\label{qedd}
E_n=2m_e-m_e\frac{\alpha^2}{4n^2}.
\ee

\begin{table}[htbp]
  \centering
  \begin{tabular}{|c|c|c|}
\hline
n & t-2 &  $-\frac{\alpha^2}{4n^2}$ \\
\hline

  1  & $-1.33128\times 10^{-5}$ &  $-1.33128\times 10^{-5}$   \\
  \hline
  2  & $-3.32830\times 10^{-6}$ &  $-3.32821\times 10^{-6}$    \\
  \hline
  3  & $-1.47926\times 10^{-6}$ &  $-1.47920\times 10^{-6}$     \\
  \hline
  4  & $-8.32084\times 10^{-7}$ &  $-8.32052\times 10^{-7}$    \\
  \hline
  5  & $-5.32534\times 10^{-7}$  &  $-5.32514\times 10^{-7}$      \\

  \hline
  \end{tabular}
  \caption{\label{tabi1} Comparing with the known results}
  \end{table}

 \begin{figure}[h]
\center
\includegraphics[width=0.4\textwidth]{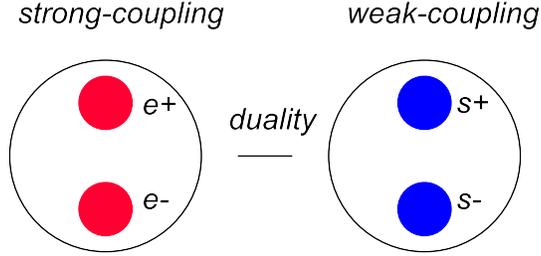}
\caption{\label{electric} The soliton contribution of positronium ($e^+e^-$) bound state in weak-coupling can been equivalent to the system of ($e^+e^-$) in strong-couping.  }
\end{figure}
 We discuss the physical meaning of non-perturbative term. The leading term of equation (\ref{qede}) is
\bea
&1&-4n\alpha i \int_{\widetilde{x}_1}^{\widetilde{x}_2} dx x(1-x)\nonumber\\
=&1&-\frac{2n\alpha}{3}\sqrt{\frac{4\Gamma(\alpha,n,a)}{t^2}-1}(1+\frac{2\Gamma(\alpha,n,a)}{t^2})\nonumber\\
\approx &1&-\frac{2n\alpha}{3}\sqrt{\frac{4\Omega(n\alpha,a)}{t^2}-1}(1+\frac{2\Omega(n\alpha,a)}{t^2})\nonumber\\
=&1&-\frac{2an}{3\alpha}\sqrt{\frac{4}{t^2}-1}(1+\frac{2}{t^2})=0.\nonumber
\eea
 We can view the $n\alpha$ as the effective coupling constant $\widetilde{\alpha}$. The leading term of $\frac{\widetilde{m}_e^2(\alpha,n,a)}{m_e^2}$ (equation (\ref{mm})) that is $\Omega(n\alpha,a)=\Omega(\widetilde{\alpha},a)$ depends on the effective coupling constant $\widetilde{\alpha}$. The soliton contribution of positronium ($e^+e^-$) bound state in weak-coupling can been equivalent to the system of ($e^+e^-$) in strong-coupling (FIG. \ref{electric}). This is also a kind of weak-strong transformation similar to electric-magnetic duality\cite{Dirac:1931kp,Montonen:1977sn}.

%%%%%
%%%%%
%%%%%
%%%%%
%%%%%

\section{Vector meson spectrum in QCD}
Finally, we discuss the vector meson spectrum in QCD. The gluon-gluon interactions in QCD have no analogue in QED, and it can be shown that they lead to properties of the strong interaction that differ markedly from those of the electromagnetic interaction. These properties are colour confinement and asymptotic freedom. The colour confinement have many theories, e.g. the dual superconductor picture of confinement\cite{Nielsen:1979xu}.
We discuss the vector meson spectrum which is similar to the positronium.
\begin{figure}[h]
\center
\includegraphics[width=0.6\textwidth]{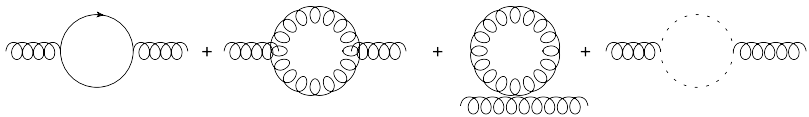}
\caption{\label{QCD} The gluon propagator in order $g^2$.  }
\end{figure}
The $i\Pi^{\mu a,\nu b}_2(q)$ has the following structure
\bea
i\Pi^{\mu a,\nu b}_2(q)=i(q^2g^{\mu\nu}-q^{\mu}q^{\nu})\delta^{ab}\Pi_2(q^2).\nonumber
\eea
Where $\Pi_2(q^2)$ is calculated from FIG. \ref{QCD} \cite{Peskin:1995ev}

\bea
\label{QCDaa}
\Pi_2(q^2)=\Pi_2^f(q^2)+\Pi_2^g(q^2).
\eea
Where the $\Pi_2^f(q^2)$ is the fermion loop diagram contribution and $\Pi_2^g(q^2)$ is the three diagrams of FIG. \ref{QCD} from the pure gauge sector.
\bea
&&\Pi_2^f(q^2)=\frac{-2C(r)n_f\alpha_g}{\pi}\int_{C[0,1]}dxx(1-x)\frac{\Gamma(2-\frac{d}{2})}
{(m_q^2-x(1-x)q^2)^{2-\frac{d}{2}}}\nonumber\\
&&\Pi_2^g(q^2)=\frac{C_2(G)\alpha_g}{4\pi}\int_{C[0,1]}dx\frac{\Gamma(2-\frac{d}{2})}{(-x(1-x)q^2)^{2-\frac{d}{2}}}
[(1-\frac{d}{2})(1-2x)^2+2].
\eea

Here $\alpha_g$ is $\alpha_g=\frac{g^2}{4\pi}$, the $n_f$ is the number of fermion species, $C(r)$ is $C(r)=\frac{1}{2}$ for fundamental representations, $C_2(G)$ is $C_2(G)=N$ for $SU(N)$ and the $m_q$ is the free quark mass. We then consider the soliton contribution. Similar to QED, we first present soliton contributions $\Pi_2^{f,s}(q^2)$
\bea
&&\Pi_2^{f,s}(q^2)=\frac{2C(r)n_f\alpha_g}{\pi}\int_{C[0,1]}dxx(1-x)\frac{\Gamma(2-\frac{d}{2})}
{(\widetilde{m}_q^2-x(1-x)q^2)^{2-\frac{d}{2}}}\nonumber,
\eea
which comes from the dual field of the fermion. When come to the dual soliton of the gluon, we find the different case. In non-Abelian gauge theories, the Pauli-Villars regularization does
not work, which is a gauge invariant regularization but not a gauge covariant regularization. To preserve the gauge symmetry, we
don't have the soliton contribution corresponding to gluon (or the mass of soliton being infinite). This leads to the colour confinement. Then the energy eigenvalues of the bound states are solutions of equation
\bea
\label{QC}
1-\Pi_2^f(q^2)-\Pi_2^g(q^2)-\Pi_2^{f,s}(q^2)=0.
\eea

\begin{figure}[h]
\center
\includegraphics[width=0.4\textwidth]{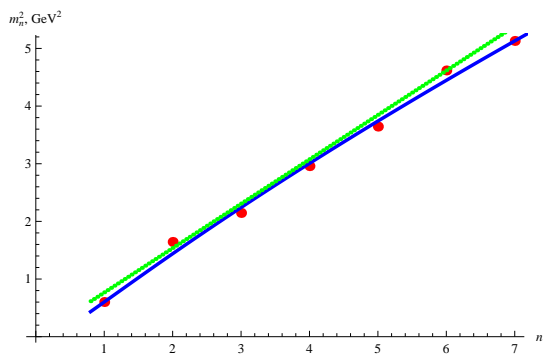}
\caption{\label{QCDE} The red dots denote the experiment data of the squared masses of $\rho$ resonances. The blue curve is the fit $m^2(n)\sim (1.6315578086364482n^{1/3}-0.8560578086364481)^2{\rm (GeV^2)}$. The green curve is the fit $m^2(n)\sim 0.77n{\rm (GeV^2)}$.}
\end{figure}

The leading contribution of equation (\ref{QC}) is
\bea \label{QC1}
&&C(r)n_f\frac{2nf(\alpha_g)}{3}\sqrt{\frac{4}{t^2}-1}(1+\frac{2}{t^2})-h_q\Gamma(2-\frac{d}{2})\nonumber\\
&&=a_qn\sqrt{\frac{4}{t^2}-1}(1+\frac{2}{t^2})-b_q=0 .
\eea
The concrete expression of the function $f(\alpha_g)$ is unimportant for our discussion. The constant $a_q$ and $b_q$ are defined as $a_q=C(r)n_f\frac{2f(\alpha_g)}{3}$ and $b_q=h_q\Gamma(2-\frac{d}{2})\rightarrow \infty$. We obtain the solution of equation (\ref{QC1}) which is
\bea
t^2(n)\approx (\frac{4na_q}{b_q})^{2/3} \quad\ \Rightarrow \quad\ m(n)\approx m_q(\frac{4na_q}{b_q})^{1/3}.
\eea
The observed physical mass $m_{{\rm th}}(n)$ is $m_{{\rm th}}(n)=m(n)-c_q=m_q(\frac{4na_q}{b_q})^{1/3}-c_q$, where the $c_q$ and $m_q(\frac{4a_q}{b_q})^{1/3}$ can determined by the experimental data fitting. We obtain $c_q=856.0578086364481{\rm (GeV)}$ and $m_q(\frac{4a_q}{b_q})^{1/3}=1631.5578086364482(GeV)$ which is consistent with the experimental data \cite{Eidelman:2004wy} (Table \ref{mesonmass}, FIG. \ref{QCDE}). It's difficult to find free quark because the $t$ is very small, which leads to the colour confinement.

\begin{table}[htbp]
  \centering
  \begin{tabular}{|c|c|c|c|c|c|c|c|}
\hline
  $\rho$&1&2&3&4&5&6&7  \\
\hline
  $m_{{\rm ex}}$(GeV) &0.7755&1.282&1.465&1.720&1.909&2.149&2.265  \\
  \hline
  $m_{{\rm th}}$(GeV) &0.7755&1.200&1.497&1.733&1.934&2.109&2.265  \\
  \hline
  error                &0$\%$&6$\%$&2$\%$&0.7$\%$&1.3$\%$&1.9$\%$&0$\%$  \\
\hline
  \end{tabular}
  \caption{\label{mesonmass} The theoretical and experimental values for vector meson $\rho$ masses. $m_{{\rm th}}(n)\sim (1631.5578086364482n^{1/3}-856.0578086364481){\rm (GeV)}$. }
  \end{table}

To compare with the Regge theory, we also give the picture of $t^2(n)\sim 0.77n{\rm (GeV^2)}$ which is known as the Regge behavior \cite{Collins:1977jy} in FIG. \ref{QCDE}. Our new experimental data fitting  is as good as Regge theory.
%%%%%
%%%%%
%%%

\section{Conclusions and Discussions }
In this paper, we give a new method to calculate the the excited spectrum of bound states in Quantum Field Theory. There is no correctly bound states pole in the two-point function at one loop in traditional method. We need to take into account the soliton contribution which can regulate the ultraviolet divergence. We discuss the bound state of massive Thirring model, the positronium ($e^+e^-$) in QED and the vector meson in QCD separately. We also give a new way to obtain the mass of soliton from the stationary condition (gap equation). Our results agree with experimental data to high precision. We argue that the hypothetic $X_{17}$ particle in decay of $\rm{{}^8 Be}$ and $\rm{{}^4 He}$ is a soliton. Especially for vector meson, we find squared masses of $\rho$ resonances is $m^2(n)\sim (an^{1/3}-b)^2$. The new experimental data fitting  is as good as Regge theory. To make the results accurate, we need to calculate the higher order loops.

%%%%%
%%%%%
%%%%%
%%%%%
%%%%%

\section*{Acknowledgments}
This work is supported by Chinese Universities Scientific Fund Grant No. 2452018158. We would like to thank Dr. Wei He, Youwei Li and Suzhi Wu for helpful discussions.

%%%%%%%%%%%%%%%%%%%%%%%%%%%%%%%%%%%%%%%%%%%%%%%%%%%%

\end{document}